\documentclass[12pt]{article}
\usepackage{enumerate}
\usepackage{amsfonts}
\usepackage{amsmath}
\usepackage{amssymb}
\usepackage{amsthm}
\usepackage{latexsym}
\usepackage{color}

\def\H{\mathcal{H}}

\def\T{\mathfrak{T}}
\def\B{\mathfrak{B}}

\newcommand{\supp}{\mathrm{supp}}
\newcommand{\rank}{\mathrm{rank}}
\newcommand{\id}{\mathrm{Id}}
\newcommand{\Tr}{\mathrm{Tr}}

\newcommand{\djn}{\mathrm{dj}}
\newcommand{\shs}{\hspace{1pt}}
\newcounter{defin}  \newcounter{lemma}  \newcounter{theorem}
\newcounter{property} \newcounter{corol}  \newcounter{remark} \newcounter{example}
\newenvironment{lemma}{\par\refstepcounter{lemma}     \textbf{Lemma \thelemma.} }{\rm\par}
\newenvironment{theorem}{\par\refstepcounter{theorem}     \textbf{Theorem \thetheorem.}\ }{\rm\par}

\newenvironment{definition}{\par\refstepcounter{defin}     \textbf{Definition \thedefin.}\ }{\rm\par}

\begin{document}


\title{New monotonicity property of the quantum relative entropy}


\author{M.E. Shirokov\footnote{email:msh@mi.ras.ru}\\Steklov Mathematical Institute, Moscow}
\date{}
\maketitle



\begin{abstract} It is proved that the local discontinuity jumps  of the quantum relative entropy
do not increase under action of quantum channels and operations.
\end{abstract}



\section{Introduction}

The quantum relative entropy is one of the basic characteristics of quantum states, which is used essentially in
the study of  information and statistical properties of quantum systems
and channels  \cite{IRE-1,IRE-2,O&P,W}.\smallskip

From the mathematical point of view,  the quantum relative entropy $D(\rho\|\shs\sigma)$ is a lower semicontinuous jointly convex function of a pair $(\rho,\sigma)$ of quantum
states (or, more generally, positive trace class operators) taking values in $[0,+\infty]$. One of the fundamental properties of the quantum relative entropy is its monotonicity
under actions of quantum  operations (completely positive
trace-non-increasing linear maps), which means that
\begin{equation}\label{m-prop+}
D(\Phi(\rho)\|\shs \Phi(\sigma))\leq D(\rho\shs\|\shs\sigma)
\end{equation}
for an arbitrary  quantum operation $\Phi$ and any quantum states $\rho$ and $\sigma$ (with possible value $+\infty$ in one or both sides) \cite{L-REM}.
Recently, it was shown that this property is valid for any trace-non-increasing positive linear maps $\Phi$ \cite{Reed}.\smallskip

The (joint) lower semicontinuity of the quantum relative entropy means that
\begin{equation}\label{D-l-s-c+}
\liminf_{n\to+\infty}D(\rho_n\|\shs\sigma_n)\geq D(\rho_0\|\shs\sigma_0)
\end{equation}
for any sequences  $\,\{\rho_n\}$ and $\{\sigma_n\}$ of quantum states converging, respectively,
to states $\rho_0$ and $\sigma_0$ (with possible value $+\infty$ in one or both sides) \cite{L-2,O&P,W}. So, if
$D(\rho_0\|\shs\sigma_0)$ is finite then the quantity
\begin{equation}\label{D-l-s-c+}
\djn (\{D(\rho_n\|\shs\sigma_n)\})\doteq \limsup_{n\to+\infty}D(\rho_n\|\shs\sigma_n)-D(\rho_0\|\shs\sigma_0)
\end{equation}
is  a nonnegative number or $+\infty$. It can be called the (maximal) jump of the quantum relative entropy
under passing to the limit $n\to+\infty$ (corresponding to the converging sequences  $\,\{\rho_n\}$ and $\{\sigma_n\}$).
By the lower semicontinuity of the quantum relative entropy we have
\begin{equation*}
 \djn (\{D(\rho_n\|\shs\sigma_n)\})=0\quad\Leftrightarrow\quad \exists\lim_{n\to+\infty}D(\rho_n\|\shs\sigma_n)=D(\rho_0\|\shs\sigma_0).
\end{equation*}

In this article we prove that the quantity $\djn(\{D(\rho_n\|\shs\sigma_n)\})$ does not increase under action of an arbitrary quantum operation $\Phi$, i.e.
\begin{equation}\label{nmp}
  \djn (\{D(\Phi(\rho_n)\|\shs\Phi(\sigma_n)\})\leq \djn (\{D(\rho_n\|\shs\sigma_n)\})
\end{equation}
for any sequences  $\,\{\rho_n\}$ and $\{\sigma_n\}$ of quantum states converging, respectively,
to states $\rho_0$ and $\sigma_0$ such that $D(\rho_0\|\shs\sigma_0)$ is finite. The quantity
in the l.h.s. of (\ref{nmp}) is well defined, since
the finiteness of $D(\rho_0\|\shs\sigma_0)$ implies the finiteness of $D(\Phi(\rho_0)\|\shs\Phi(\sigma_0))$ by the monotonicity property (\ref{m-prop+}).

The property (\ref{nmp}) strengthens the claim of Theorem 2 in \cite{REC} about preserving local continuity of the quantum relative entropy by quantum operations, since
it means (in our notation) that
\begin{equation*}
 \djn (\{D(\rho_n\|\shs\sigma_n)\})=0\quad\Rightarrow\quad \djn (\{D(\Phi(\rho_n)\|\shs\Phi(\sigma_n))\})=0.
\end{equation*}
Note that the proof of property (\ref{nmp}) presented below is completely different from (and more simple than) the proof
of Theorem 2 in \cite{REC}.

\section{Preliminaries}

\subsection{Basic notations}

Let $\mathcal{H}$ be a separable Hilbert space,
$\mathfrak{B}(\mathcal{H})$ the algebra of all bounded operators on $\mathcal{H}$ with the operator norm $\|\cdot\|$ and $\mathfrak{T}( \mathcal{H})$ the
Banach space of all trace-class
operators on $\mathcal{H}$  with the trace norm $\|\!\cdot\!\|_1$. Let
$\mathfrak{S}(\mathcal{H})$ be  the set of quantum states (positive operators
in $\mathfrak{T}(\mathcal{H})$ with unit trace) \cite{H-SCI,Wilde,BSimon}.

Denote by $I_{\mathcal{H}}$ the unit operator on a Hilbert space
$\mathcal{H}$ and by $\id_{\mathcal{\H}}$ the identity
transformation of the Banach space $\mathfrak{T}(\mathcal{H})$.\smallskip

The \emph{von Neumann entropy} of a quantum state
$\rho \in \mathfrak{S}(\H)$ is  defined by the formula
$S(\rho)=\Tr\eta(\rho)$, where  $\eta(x)=-x\ln x$ if $x>0$
and $\eta(0)=0$. It is a concave lower semicontinuous function on the set~$\mathfrak{S}(\H)$ taking values in~$[0,+\infty]$ \cite{W,H-SCI,L-2}.
\smallskip

We will use the  homogeneous extension of the von Neumann entropy to the positive cone $\T_+(\H)$ defined as
\begin{equation}\label{S-ext}
S(\rho)\doteq(\Tr\rho)S(\rho/\Tr\rho)=\Tr\eta(\rho)-\eta(\Tr\rho)
\end{equation}
for any nonzero operator $\rho$ in $\T_+(\H)$ and equal to $0$ at the zero operator \cite{L-2}.\smallskip

Let $H$ be a positive (semi-definite)  operator on a Hilbert space $\mathcal{H}$ (we will always assume that positive operators are self-adjoint). Denote by $\mathcal{D}(H)$ the domain of $H$. For any positive operator $\rho\in\T(\H)$ we will define the quantity $\Tr H\rho$ by the rule
\begin{equation}\label{H-fun}
\Tr H\rho=
\left\{\begin{array}{l}
        \sup_n \Tr P_n H\rho\;\; \textrm{if}\;\;  \supp\rho\subseteq {\rm cl}(\mathcal{D}(H))\\
        +\infty\;\;\textrm{otherwise,}
        \end{array}\right.
\end{equation}
where $P_n$ is the spectral projector of $H$ corresponding to the interval $[0,n]$, ${\rm cl}(\mathcal{D}(H))$ is the closure of $\mathcal{D}(H)$ and $\mathrm{supp}\rho$ is the support of the operator $\rho$ -- the closed subspace spanned by the eigenvectors of $\rho$ corresponding to its positive eigenvalues. \smallskip

We will use the following notion introduced in \cite{REC}. \smallskip

\begin{definition}\label{scs-def} A double sequence $\{P^n_m\}_{n\geq0,m\geq m_0}$ ($m_0\in\mathbb{N}$) of finite rank  projectors is \emph{completely consistent}
with a sequence $\{\sigma_n\}\subset\T_+(\H)$ converging to an  operator $\sigma_0$ if
\begin{equation}\label{P-prop}
 P^n_{m}\leq P^n_{m+1}, \qquad \bigvee_{m\geq m_0}P^n_m\geq Q_n,
\end{equation}
where $Q_n$ is the projector onto the support of $\sigma_n$, and
\begin{equation}\label{P-prop+}
P^n_m\sigma_n=\sigma_nP^n_m,\qquad  \rank P^n_m\sigma_n=\rank P^n_m,\qquad \|\cdot\|\,\textrm{-}\!\!\lim_{n\to+\infty}P^n_m=P^0_m
\end{equation}
for all $m\geq m_0$ and $n\geq0$, where the limit in the operator norm topology.\smallskip
\end{definition}

It is essential that  \emph{for any sequence $\{\sigma_n\}\subset\T_+(\H)$ converging to an  operator $\sigma_0$
there exists a double sequence $\{P^n_m\}_{n\geq0,m\geq m_0}$  of finite rank  projectors  completely consistent
with the sequence $\{\sigma_n\}$} \cite[Lemma 4]{REC}.
\smallskip

For a lower semicontinuous function $f$ on a metric space $X$ and a given
sequence $\{x_n\}\subset X$ converging to a point $x_0\in X$ such that $f(x_0)<+\infty$ we will use the quantity
\begin{equation}\label{dj}
\djn(\{f(x_n)\})\doteq\limsup_{n\to+\infty}f(x_n)-f(x_0).
\end{equation}
The lower semicontinuity of $f$ implies that $\djn(\{f(x_n)\})\geq0$ and that
$$
\djn(\{f(x_n)\})\geq0\quad\Leftrightarrow\quad \exists\lim_{n\to+\infty}f(x_n)=f(x_0).
$$

We will also use  the strengthened  version of Dini's lemma.\smallskip

\begin{lemma}\label{Dini+} \emph{Let $\{a_n\}_{n\geq0}$ be a sequence of nonnegative numbers such that
$$
0\leq\Delta\doteq\limsup_{n\to+\infty}a_n-a_0<+\infty.
$$
Let $\{a_m^n\}_{n\geq0,m>m_0}$ be a double sequence  such that $a_m^n\leq a_{m+1}^n$ for all $n\geq0$ and $m\geq m_0$,
$$
\lim_{m\to+\infty}a^m_n= a_n \quad  \forall n\geq 0 \qquad \textrm{and} \qquad \liminf_{n\to+\infty}a^m_n\geq a^{m}_0\quad  \forall m\geq m_0.
$$
Then}
$$
\lim_{m\to+\infty}\sup_{n\geq0}|a_n-a^m_n|\leq \Delta.
$$
\end{lemma}

\emph{Proof.} Let $\varepsilon>0$ be arbitrary. The assumptions of the lemma imply the existence of $n^1_{\varepsilon}>0$
and $m^1_{\varepsilon}>m_0$ such that
$a_n\leq a_0+\Delta+\varepsilon$ for all $n\geq n^1_{\varepsilon}$ and $a^m_0\geq a_0-\varepsilon$ for all $m\geq m^1_{\varepsilon}$.
Since $\liminf_{n\to+\infty}a^{m^1_{\varepsilon}}_n\geq a^{m^1_{\varepsilon}}_0$, there is $n^2_{\varepsilon}$
such that $a^{m^1_{\varepsilon}}_n\geq a^{m^1_{\varepsilon}}_0-\varepsilon$ for all $n\geq n^2_{\varepsilon}$. Hence,
$$
a^{m}_n\geq a^{m^1_{\varepsilon}}_n\geq a_n-\Delta-3\varepsilon\qquad \forall n\geq n_{\varepsilon}\doteq\max\{n^1_{\varepsilon},n^2_{\varepsilon}\},\quad \forall m\geq m^1_{\varepsilon}.
$$
Since $\lim_{m\to+\infty}a^m_n= a_n$ for all $n$, there is $m^2_{\varepsilon}$ such that $a^{m}_n\geq a_n-3\varepsilon$
for all $n< n_\varepsilon$ provided that $m\geq m^2_{\varepsilon}$. Thus $\sup_{n\geq0}|a_n-a^m_n|\leq \Delta+3\varepsilon$ for all
$m\geq \max\{m^1_{\varepsilon},m^2_{\varepsilon}\}$. $\Box$

\subsection{Lindblad's extension of the quantum relative entropy}

The \emph{quantum relative entropy} for two states $\rho$ and
$\sigma$ in $\mathfrak{S}(\mathcal{H})$ is defined as
\begin{equation*}
D(\rho\shs\|\shs\sigma)=\sum_i\langle
\varphi_i|\,\rho\ln\rho-\rho\ln\sigma\,|\varphi_i\rangle,
\end{equation*}
where $\{\varphi_i\}$ is the orthonormal basis of
eigenvectors of the state $\rho$ and it is assumed that
$D(\rho\,\|\sigma)=+\infty$ if $\,\mathrm{supp}\rho\shs$ is not
contained in $\shs\mathrm{supp}\shs\sigma$ \cite{Ume,W,L-2}.\smallskip

We will use Lindblad's extension of the quantum relative entropy to any positive
operators $\rho$ and
$\sigma$ in $\mathfrak{T}(\mathcal{H})$ defined as
\begin{equation*}
D(\rho\shs\|\shs\sigma)=\sum_i\langle\varphi_i|\,\rho\ln\rho-\rho\ln\sigma\,|\varphi_i\rangle+\Tr\sigma-\Tr\rho,
\end{equation*}
where $\{\varphi_i\}$ is the orthonormal basis of
eigenvectors of the operator  $\rho$ and it is assumed that $\,D(0\|\shs\sigma)=\Tr\sigma\,$ and
$\,D(\rho\shs\|\sigma)=+\infty\,$ if $\,\mathrm{supp}\rho\shs$ is not
contained in $\shs\mathrm{supp}\shs\sigma$ (in particular, if $\rho\neq0$ and $\sigma=0$)
\cite{L-2}.

The function $(\rho,\sigma)\mapsto D(\rho\shs\|\shs\sigma)$ is nonnegative lower semicontinuous and jointly convex on
$\T_+(\H)\times\T_+(\H)$. We will use the following properties of this function:
\begin{itemize}
  \item for any $\rho,\sigma\in\T_+(\H)$ and $c\geq0$ the following equalities  hold:
  \begin{equation}\label{D-mul}
  D(c\rho\shs\|\shs c\sigma)=cD(\rho\shs\|\shs \sigma),\qquad\qquad\qquad\qquad\quad\;
  \end{equation}
\begin{equation}\label{D-c-id}
D(\rho\shs\|\shs c\sigma)=D(\rho\shs\|\shs\sigma)-\Tr\rho\ln c+(c-1)\Tr\sigma;
\end{equation}
  \item for any $\rho,\sigma,\omega$ and $\vartheta$ in $\T_+(\H)$ the following inequality holds
  \begin{equation}\label{D-sum-g}
    D(\rho+\sigma\shs\|\shs \omega+\vartheta)\leq D(\rho\shs\|\shs \omega)+D(\sigma\shs\|\shs \vartheta),
  \end{equation}
  if $\rho\sigma=\rho\vartheta=\sigma\omega=\omega\vartheta=0$ then
  \begin{equation}\label{D-sum}
    D(\rho+\sigma\shs\|\shs \omega+\vartheta)=D(\rho\shs\|\shs \omega)+D(\sigma\shs\|\shs \vartheta).
  \end{equation}
 \end{itemize}

Inequality (\ref{D-sum-g}) is a direct corollary of the joint convexity of the relative entropy and identity (\ref{D-mul}). Equality (\ref{D-sum}) follows from the definition \cite{L-2}.

If the extended von Neumann entropy $S(\rho)$ of $\rho$ (defined in (\ref{S-ext})) is finite
then
\begin{equation}\label{re-exp}
D(\rho\shs\|\shs\sigma)=\Tr\rho(-\ln\sigma)-S(\rho)-\eta(\Tr\rho)+\Tr\sigma-\Tr\rho,
\end{equation}
where $\Tr\rho(-\ln\sigma)$ is defined according to the rule (\ref{H-fun}).

We will use Donald's identity
\begin{equation}\label{Donald}
pD(\rho\|\omega)+\bar{p}D(\sigma\|\omega)=pD(\rho\|p\rho+\bar{p}\sigma)+\bar{p}D(\sigma\|p\rho+\bar{p}\sigma)+D(p\rho+\bar{p}\sigma\|\omega)
\end{equation}
where $\bar{p}=1-p$, valid for arbitrary operators $\rho$, $\sigma$ and $\omega$ in $\T_+(\H)$ and any $p\in[0,1]$ \cite{Donald}.\footnote{In Lemma 2 in \cite{Donald}
it was assumed that $\rho$, $\sigma$ and $\omega$ are (normal) states. The generalization to arbitrary operators in $\T_+(\H)$
can be done by using identities (\ref{D-mul}) and (\ref{D-c-id}).}

\section{The main result}

A basis property of the quantum relative entropy is its monotonicity under quantum operations (completely positive
trace-non-increasing linear maps), which means that
\begin{equation}\label{m-prop}
D(\Phi(\rho)\|\shs \Phi(\sigma))\leq D(\rho\shs\|\shs\sigma)
\end{equation}
for an arbitrary  quantum operation $\Phi:\T(\H_A)\to\T(\H_B)$ and any operators $\rho$ and $\sigma$ in $\T_+(\H_A)$ \cite{L-REM}.\footnote{Here and in what follows $D(\cdot\shs\|\shs\cdot)$ is Lindblad's extension of the Umegaki relative entropy to operators in $\T_+(\H_A)$ described in Section 2.2.}\smallskip

It turns out that a similar monotonicity property holds for the local discontinuity jumps of the quantum relative entropy.

\smallskip

\begin{theorem}\label{conv-crit} \emph{Let $\,\{\rho_n\}$ and $\{\sigma_n\}$ be sequences of operators in $\,\T_+(\H_A)$ converging, respectively,
to operators  $\rho_0$ and $\sigma_0$ such that $D(\rho_0\|\shs\sigma_0)<+\infty$. Then\smallskip
\begin{equation}\label{D-cont}
\limsup_{n\to+\infty}D(\Phi(\rho_n)\|\shs\Phi(\sigma_n))-D(\Phi(\rho_0)\|\shs\Phi(\sigma_0))\leq\limsup_{n\to+\infty}D(\rho_n\|\shs\sigma_n)-D(\rho_0\|\shs\sigma_0)
\end{equation}
for any quantum operation $\Phi:\T(\H_A)\to\T(\H_B)$.}\smallskip
\end{theorem}

\emph{Proof.} Assume first that $\Phi$ is a quantum channel having the Stinespring representation
\begin{equation}\label{St-rep}
\Phi(\rho)=\mathrm{Tr}_E V\rho V^*,\quad \rho\in\T(\H_A),
\end{equation}
where $V$ is an isometry from $\,\mathcal{H}_{A}$ to $\mathcal{H}_{BE}$ \cite{H-SCI,Wilde}.

If $\sigma_0=0$ then $\rho_0=0$ (otherwise $D(\rho_0\|\shs\sigma_0)=+\infty$) and (\ref{D-cont}) follows directly from the
monotonicity property (\ref{m-prop}). So, we will assume that $\sigma_0\neq0$ and, hence, $\Phi(\sigma_0)\neq0$. 

It suffices to consider the case when $D(\rho_n\|\shs\sigma_n)<+\infty$ for all $n$ and\footnote{We will use the notation (\ref{dj}).}
$$
\Delta=\djn(\{D(\rho_n\|\shs\sigma_n)\})\doteq\limsup_{n\to+\infty}D(\rho_n\|\shs\sigma_n)-D(\rho_0\|\shs\sigma_0)<+\infty.
$$

Since $V$ is an isometry, we have
\begin{equation}\label{a-n}
a_n\doteq D(V\rho_nV^*\|\shs V\sigma_nV^*)=D(\rho_n\|\shs\sigma_n),\;\, \forall n\geq0.
\end{equation}
Let $\{P^n_m\}_{n\geq0,m\geq m_0}$ be a double sequence of finite rank projectors in $\B(\H_B)$
completely consistent  with the sequence $\{\Phi(\sigma_n)\}$ (Definition \ref{scs-def} in Section 2.1) which exists by Lemma 4 in \cite{REC}. Consider the double sequence
$$
a_n^m=D((P^n_m\otimes I_E)V\rho_nV^*(P^n_m\otimes I_E)\|\shs (P^n_m\otimes I_E)V\sigma_nV^*(P^n_m\otimes I_E)),\quad n\geq0,m\geq m_0
$$
of nonnegative numbers. By Lemma 4 in \cite{L-2} the  conditions in (\ref{P-prop}) imply that
$$
a_n^m\leq a_n^{m+1}\quad\textrm{and}\quad
\lim_{m\to+\infty}a^m_n= a_n \quad  \forall n\geq 0.
$$
The third condition in (\ref{P-prop+}) shows that
$$
\lim_{n\to+\infty}(P^n_m\otimes I_E)V\omega_nV^*(P^n_m\otimes I_E)=(P^0_m\otimes I_E)V\omega_0V^*(P^0_m\otimes I_E),\quad \omega=\rho,\sigma,
$$
for any $m\geq m_0$. So, by the lower semicontinuity of the quantum relative entropy we have
$$
\liminf_{n\to+\infty}a^m_n\geq a^{m}_0\quad  \forall m\geq m_0.
$$
Thus, by using Lemma \ref{Dini+} in Section 2.1 we obtain
\begin{equation}\label{est-a+}
\lim_{m\to+\infty}\sup_{n\geq0}|a_n-a^m_n|\leq\Delta.
\end{equation}
Lemma 3  in \cite{L-2} implies that
\begin{equation*}
  D((\bar{P}^n_m\otimes I_E)V\rho_nV^*(\bar{P}^n_m\otimes I_E)\|\shs (\bar{P}^n_m\otimes I_E)V\sigma_nV^*(\bar{P}^n_m\otimes I_E))\leq a_n-a^m_n
\end{equation*}
for all $n\geq0,m\geq m_0$, where $\bar{P}^n_m=I_B-P^n_m$. So, it follows from (\ref{est-a+}) that for any $\varepsilon>0$ there is $m_\varepsilon$
such that
\begin{equation*}
\sup_{n\geq0}  D((\bar{P}^n_m\otimes I_E)V\rho_nV^*(\bar{P}^n_m\otimes I_E)\|\shs (\bar{P}^n_m\otimes I_E)V\sigma_nV^*(\bar{P}^n_m\otimes I_E))\leq \Delta+\varepsilon
\end{equation*}
for all $m\geq m_\varepsilon$. Since 
$$
\Tr_E(\bar{P}^n_m\otimes I_E)V\omega_nV^*(\bar{P}^n_m\otimes I_E)=\bar{P}^n_m\Phi(\omega_n)\bar{P}^n_m,\quad \omega=\rho,\sigma,
$$
by monotonicity property (\ref{m-prop}) of the quantum relative entropy we have
\begin{equation}\label{est-a}
\djn(\{D(\bar{P}^n_m\Phi(\rho_n)\bar{P}^n_m\|\shs \bar{P}^n_m\Phi(\sigma_n))\}_n)\leq  \sup_{n\geq0}D(\bar{P}^n_m\Phi(\rho_n)\bar{P}^n_m\|\shs \bar{P}^n_m\Phi(\sigma_n))\leq\Delta+\varepsilon
\end{equation}
for all $m\geq m_\varepsilon$.

By Lemma \ref{b-lemma} below the properties of the double sequence $\{P^n_m\}_{n\geq0,m\geq m_0}$ imply that
\begin{equation}\label{est-b}
\lim_{n\to+\infty}D(P^n_m\Phi(\rho_n)P^n_m\|\shs P^n_m\Phi(\sigma_n))=D(P^0_m\Phi(\rho_0)P^0_m\|\shs P^0_m\Phi(\sigma_0))<+\infty
\end{equation}
for all $m\geq m_0$. Hence, it follows from (\ref{est-a}) and (\ref{est-b}) that
\begin{equation}\label{est-d}
\djn(\{D(P^n_m\Phi(\rho_n)P^n_m\|\shs P^n_m\Phi(\sigma_n))+D(\bar{P}^n_m\Phi(\rho_n)\bar{P}^n_m\|\shs \bar{P}^n_m\Phi(\sigma_n))\}_n)\leq\Delta+\varepsilon
\end{equation}
for all $m\geq m_\varepsilon$. Since identity (\ref{D-sum}) shows that
\begin{equation}\label{est-e}
\begin{array}{c}
D(P^n_m\Phi(\rho_n)P^n_m\|\shs P^n_m\Phi(\sigma_n))+D(\bar{P}^n_m\Phi(\rho_n)\bar{P}^n_m\|\shs \bar{P}^n_m\Phi(\sigma_n))\\\\
=D(P^n_m\Phi(\rho_n)P^n_m+\bar{P}^n_m\Phi(\rho_n)\bar{P}^n_m\|\shs \Phi(\sigma_n))\\\\=D(\frac{1}{2}(\Phi(\rho_n)+U^n_m\Phi(\rho_n)[U^n_m]^*)\|\shs \Phi(\sigma_n)),
\end{array}
\end{equation}
where $U^n_m=2P^n_m-I_B$ is a unitary operator for all $n\geq0,m\geq m_0$, it follows from (\ref{est-d}) that
\begin{equation}\label{est-f}
\djn(\{D(\textstyle\frac{1}{2}(\Phi(\rho_n)+U^n_m\Phi(\rho_n)[U^n_m]^*)\|\shs \Phi(\sigma_n))\}_n)\leq\Delta+\varepsilon\quad \forall m\geq m_\varepsilon.
\end{equation}

Since $U^n_m\Phi(\sigma_n)[U^n_m]^*=\Phi(\sigma_n)$ for all $m$ and $n$ by the first property in (\ref{P-prop+}), by using Donald's identity (\ref{Donald}) we obtain
\begin{equation*}
\begin{array}{c}
D(\Phi(\rho_n)\|\shs \Phi(\sigma_n))=\textstyle\frac{1}{2}D(\Phi(\rho_n)\|\shs \Phi(\sigma_n))+\frac{1}{2}D(U^n_m\Phi(\rho_n)[U^n_m]^*\|\shs \Phi(\sigma_n))   \\\\=D(\textstyle\frac{1}{2}(\Phi(\rho_n)+U^n_m\Phi(\rho_n)[U^n_m]^*)\|\shs \Phi(\sigma_n))+\textstyle\frac{1}{2}(\Phi(\rho_n)\shs\|\frac{1}{2}(\Phi(\rho_n)+U^n_m\Phi(\rho_n)[U^n_m]^*))\\\\
+\frac{1}{2}D(U^n_m\Phi(\rho_n)[U^n_m]^*\shs\|\frac{1}{2}(\Phi(\rho_n)+U^n_m\Phi(\rho_n)[U^n_m]^*))\quad m\geq m_0,\, n\geq0.
\end{array}
\end{equation*}

This equality and (\ref{est-f}) shows, by Lemma \ref{b-lemma+} below, that
$$
\djn(\{D(\Phi(\rho_n)\|\shs \Phi(\sigma_n))\})=\djn(\{D(\textstyle\frac{1}{2}(\Phi(\rho_n)+U^n_m\Phi(\rho_n)[U^n_m]^*)\|\shs \Phi(\sigma_n))\}_n)\leq\Delta+\varepsilon
$$
for all $m\geq m_\varepsilon$. Since $\varepsilon$ is arbitrary, this implies (\ref{D-cont}).\smallskip

Assume that $\Phi$ is a quantum operation from $\T(\H_A)$ to $\T(\H_B)$. Consider the channel
$$
\widetilde{\Phi}_n(\rho)=\Phi_n(\rho)\oplus [\Tr(I_A-\Phi_n^*(I_B))\rho\shs]\shs\tau
$$
from $\T(\H_A)$ to $\T(\H_B\oplus\H_C)$, where $\tau$ is a pure state in a one-dimensional Hilbert space $\H_C$
and $\Phi^*_n:\B(\H_B)\to\B(\H_A)$ is the dual map to $\Phi_n$ defined by the relation $\Tr B\Phi_n(\varrho)=\Tr\Phi_n^*(B)\varrho,\,\varrho\in\T(\H_A),\,B\in \B(\H_B)$.
By identity (\ref{D-sum}) we have
$$
D(\widetilde{\Phi}(\rho_n)\|\shs \widetilde{\Phi}(\sigma_n))=D(\Phi(\rho_n)\|\shs \Phi(\sigma_n))+x_n\ln(x_n/y_n)+y_n-x_n,\quad \forall n\geq0,
$$
where $x_n=\Tr(I_A-\Phi_n^*(I_B))\rho_n$ and $y_n=\Tr(I_A-\Phi_n^*(I_B))\sigma_n$. Since $x_n\to x_0$ and $y_n\to y_0$  as $n\to+\infty$, it is clear that
$$
\djn(\{D(\Phi(\rho_n)\|\shs \Phi(\sigma_n))\})\leq\djn(\{D(\widetilde{\Phi}(\rho_n)\|\shs \widetilde{\Phi}(\sigma_n))\})\leq\djn(\{D(\rho_n\|\shs \sigma_n)\}),
$$
where the last inequality follows from the first part of the proof. $\Box$ \smallskip

\begin{lemma}\label{b-lemma} \emph{Let $\,\{\rho_n\}$ and $\{\sigma_n\}$ be sequences of operators in $\,\T_+(\H)$ converging, respectively,
to operators  $\rho_0$ and $\sigma_0\neq0$. If $\{P^n_m\}_{n\geq0,m\geq m_0}$ is a double sequence  of finite rank  projectors  completely consistent
with the sequence $\{\sigma_n\}$ (Definition 1 in Section 2) then}
\begin{equation}\label{D-cont-d+}
\lim_{n\to+\infty}\textstyle D(P^n_m\rho_nP^n_m\shs\|P^n_m\sigma_nP^n_m)=D(P^0_m\rho_0P^0_m\shs\|P^0_m\sigma_0P^0_m)<+\infty\quad \forall m\geq m_0.
\end{equation}
\end{lemma}

\emph{Proof.} Since $P^n_m\rho_n P^n_m$ tends to $P^0_m\rho_0 P^0_m$ as $\,n\to+\infty\,$ and $\,\sup_{n\geq0}\rank P^n_m\rho_n P^n_m< +\infty$  by the  conditions in (\ref{P-prop}) and (\ref{P-prop+}), by using representation (\ref{re-exp}) we see that to prove (\ref{D-cont-d+}) it suffices to show that
\begin{equation}\label{D-cont-d++}
\lim_{n\to+\infty}\|P^n_m \ln (P^n_m\sigma_n)-P^0_m \ln (P^0_m\sigma_0)\|=0
\end{equation}
for any given $m\geq m_0$. By the second condition in (\ref{P-prop+}) we have $\rank P^n_m\sigma_n=\rank P^n_m$ for
all $n\geq0$. Hence the sequence  $\{P^n_m\sigma_n+\bar{P}^n_m\}_n$
consists of bounded nondegenerate operators and converges to the nondegenerate operator $P^0_m\sigma_0+\bar{P}^0_m$
in the operator  norm by the third condition in (\ref{P-prop+}). It follows that $P^n_m\sigma_n+\bar{P}^n_m\geq\epsilon I_{\H}$ for
all $n\geq0$ and some $\epsilon>0$. So, by Proposition VIII.20 in \cite{R&S} the sequence  $\{\ln (P^n_m\sigma_n+\bar{P}^n_m)\}_n$
converges to the operator $\ln (P^0_m\sigma_0+\bar{P}^0_m)$ in the operator  norm. This implies (\ref{D-cont-d++}), since
$$
P^n_m \ln (P^n_m\sigma_n)=P^n_m\ln (P^n_m\sigma_n+\bar{P}^n_m)\quad \forall n\geq0.\;\Box
$$

\begin{lemma}\label{b-lemma+} \emph{The function
$$
(\rho,\sigma)\mapsto \textstyle D(\rho\shs\|\frac{1}{2}\rho+\frac{1}{2}\sigma)+D(\sigma\shs\|\frac{1}{2}\rho+\frac{1}{2}\sigma)
$$
is continuous on $\T_+(\H)\times\T_+(\H)$.}
\end{lemma}\smallskip

\emph{Proof.} Let $\,\{\rho_n\}$ and $\{\sigma_n\}$ be sequences of operators in $\,\T_+(\H)$ converging, respectively,
to operators  $\rho_0$ and $\sigma_0$. Since $\rho_n\leq 2(\frac{1}{2}\rho_n+\frac{1}{2}\sigma_n)$ and
$\sigma_n\leq 2(\frac{1}{2}\rho_n+\frac{1}{2}\sigma_n)$ for all $n\geq0$, Proposition 2 in \cite{DTL} implies that
$$
\lim_{n\to+\infty}\textstyle D(\rho_n\shs\|\frac{1}{2}\rho_n+\frac{1}{2}\sigma_n)=D(\rho_0\shs\|\frac{1}{2}\rho_0+\frac{1}{2}\sigma_0)<+\infty
$$
and
$$
\lim_{n\to+\infty}\textstyle D(\sigma_n\shs\|\frac{1}{2}\rho_n+\frac{1}{2}\sigma_n)=D(\sigma_0\shs\|\frac{1}{2}\rho_0+\frac{1}{2}\sigma_0)<+\infty.\,\Box
$$
\bigskip

I am grateful to A.S.Holevo, A.V.Bulinski and E.R.Loubenets for useful discussion and comments.

\end{document}